\documentclass[two column, prA]{revtex4}%
\usepackage{amsmath}
\usepackage{graphicx}
\usepackage{amsfonts}
\usepackage{amssymb}
\usepackage{color}%
\providecommand{\U}[1]{\protect\rule{.1in}{.1in}}
\ifx\pdfoutput\relax\let\pdfoutput=\undefined\fi
\newcount\msipdfoutput
\ifx\pdfoutput\undefined\else
\ifcase\pdfoutput\else
\ifx\paperwidth\undefined\else
\ifdim\paperheight=0pt\relax\else\pdfpageheight\paperheight\fi
\ifdim\paperwidth=0pt\relax\else\pdfpagewidth\paperwidth\fi
\fi\fi\fi
\begin{document}
\title{Validation of classical modeling of single-photon pulse propagation}
\author{Margaret Hawton}
\affiliation{Department of Physics, Lakehead University, Thunder Bay, ON, Canada, P7B 5E1}
\email{mhawton@lakeheadu.ca}

\begin{abstract}
"It is well-known to those who know it" that single-photon interference
experiments can be modeled classically [S. Barnett, arXiv:2207.14632 (2022)].
When a single-photon light pulse was split by a biprism good agreement with a
classical fit was obtained and the photon was counted only once, consistent
with a probabilistic interpretation [V. Jacques et al, Eur. Phys. J. D 35, 561
(2002)]. A justification for this "well know result of Quantum Optics" is
implicit in [M. Hawton, Phys. Rev A 104, 052211 (2021)] where a real covariant
field describing a single photon is first quantized. Here the theoretical
basis of this result is reviewed and the theory is extended to multiphoton
states. The crucial role of the CPT theorem in coupling to charged matter and
resolution of the photon localization problem is discussed.

\end{abstract}
\maketitle

\section{Introduction}

Since classical electromagnetic (EM) fields are real and covariant, precise
justification of their surprising success in the interpretation of
single-photon experiments \cite{BeamSplitter,Barnett,LocalPhotons} requires
quantized fields that are also real and covariant. In \cite{RealHilbert} the
real classical electromagnetic (EM) field was first quantized to give a
quantum mechanical (QM) description of one-photon states in which their state
space is augmented with a scalar product and operators describing the
momentum, energy, position and angular momentum observables.

The initial work on photon wave mechanics was based on positive energy fields
\cite{SmithRaymer}, but restriction to positive energy is inconsistent with
causal pulse propagation. According to the Hegerfeldt theorem a positive
energy field localized in a finite region for an instant spreads immediately
throughout space \cite{Hegerfeldt}. A technique that allows inclusion of
negative energy fields was devised by Mostafazadeh and co-workers
\cite{MostafazadehZamani}. They defined an operator that multiplies the
negative energy antiphoton terms by $-1$. This operator performs the same
function as the $4\times4$ matrix $\beta=\left(
\begin{array}
[c]{cc}%
\widehat{1}_{2} & 0\\
0 & -\widehat{1}_{2}%
\end{array}
\right)  $ \ in the Dirac theory of electrons and positrons. (Here
$\widehat{1}_{2}$ is a $2\times2$ unit matrix.) Number density cannot be
derived directly from a Lagrangian since it is not a conserved quantity, but
modification of the sign of the antiparticle term in the conjugate momentum
converts the conserved quantity that is generated by a phase change to photon number.

Real fields are, in fact, required for consistency with the charge-parity-time
(CPT) theorem of quantum field theory (QFT). Charge conjugation, C, exchanges
all particles with their antiparticles. The Fermion four-current is odd under
charge conjugation since electrons are exchanged with positrons. To maintain
invariance of the current-field interaction and the Dirac equation the photon
four-potential should also be odd under charge conjugation. If $A_{j}^{+}$ is
a positive energy photon four-potential and $A_{j}^{-}=A_{j}^{+\ast}$ is a
negative energy antiphoton four-potential, QFT requires that $A_{j}=\left(
A_{j}^{+}-A_{j}^{-}\right)  /\sqrt{2}i$ \cite{GellMann}.

To maintain the classical form that we seek, a covariant approach to first and
second quantization will be used here. The usual textbook choice in quantum
electrodynamics (QED) and quantum optics includes a factor $\omega_{k}^{-1/2}$
in $\widehat{A}$ where $\omega_{k}$ is the angular frequency at wave vector
$\mathbf{k}$, but here we will follow the covariant treatment in
\cite{ItzyksonZuber,VincentThesis}.

In the next Section the fields in \cite{RealHilbert} will be generalized to
include circularly polarized (CP) light for which rotation of the field
vectors mixes sine and cosine terms. For completeness descriptions of the
covariant notation, scalar product and momentum and position eigenvectors will
be included. It will be verified that only the odd field is coupled to charged
matter and localizable in a finite region. The probability amplitude to find a
photon at $\mathbf{x}$ on the $t$-hyperplane will be calculated and it will be
verified that the Born rule is satisfied. In Section III QED Fock space,
multiphoton states and the classical large photon number limit will be
discussed and in Section IV we conclude.

\section{One-photon fields, scalar product and observables}

SI units will be used. The contravariant space-time, wave vector and momentum
four-vectors are $x=x^{\mu}=\left(  ct,\mathbf{x}\right)  ,$ $k=\left(
\omega_{k}/c,\mathbf{k}\right)  $ and $p=\hbar k$ where $kx=\omega
_{k}t-\mathbf{k\cdot x}$ is invariant, the matter four-current is $J_{m}$, the
four-gradient is $\partial=\left(  \partial_{ct},-\mathbf{\nabla}\right)  $,
$\square\equiv\partial_{\mu}\partial^{\mu}=\partial_{ct}^{2}-\nabla^{2}$, the
four-potential is $A\left(  t,\mathbf{x}\right)  =A^{\mu}=\left(  \frac{\phi
}{c},\mathbf{A}\right)  $ or $a\left(  t,\mathbf{k}\right)  =\left(
a_{0},\mathbf{a}\right)  $ and $a_{\lambda}\left(  \mathbf{k}\right)  $ will
denote a Lorentz invariant scalar describing a state with definite helicity,
$\lambda$. When not written explicitly the space-time dependence of $A$ and
wavevector dependence $a$ is implied. The covariant four-vector corresponding
to $U^{\mu}=\left(  U_{0},\mathbf{U}\right)  $ is $U_{\mu}=g_{\mu\nu}U^{\nu
}=\left(  U_{0},-\mathbf{U}\right)  $ where $g_{\mu\nu}=g^{\mu\nu}$ is a
$4\times4$ diagonal matrix with diagonal $\left(  1,-1,-1,-1\right)  $. With
the mutually orthogonal polarization unit vectors $e^{\mu}$ defined such that
0 is time-like, 1 and 2 are transverse and 3 is longitudinal, $e_{0}=n^{\mu
}=\left(  1,0,0,0\right)  ,\ \mathbf{e}_{3}\left(  \mathbf{k}\right)
=\mathbf{e}_{\mathbf{k}}=\mathbf{k}/\left\vert \mathbf{k}\right\vert $ and the
definite helicity transverse unit vectors are%
\begin{equation}
\mathbf{e}_{\lambda}\left(  \mathbf{k}\right)  =\frac{1}{\sqrt{2}}\left(
\mathbf{e}_{\theta}+i\lambda\mathbf{e}_{\phi}\right)
\label{transverseeigenvectors}%
\end{equation}
for $\lambda=\pm1\ $where $\mathbf{e}_{\theta},$ $\mathbf{e}_{\phi}$ and
$\mathbf{e}_{\mathbf{k}}$ are orthonormal $\mathbf{k}$-space spherical polar
unit vectors on the $t$-hyperplane.

The four-potential describing single photons and antiphotons in position space
will be written as%
\begin{align}
A\left(  x\right)   &  =\sqrt{\frac{\hbar}{2\epsilon_{0}}}\int_{t}%
\frac{d\mathbf{k}}{\left(  2\pi\right)  ^{3}\omega_{k}}\left\{  \left[
\left(  a_{1x^{\prime}}\left(  \mathbf{k}\right)  +a_{-1x^{\prime}}\left(
\mathbf{k}\right)  \right)  \right]  \right. \nonumber\\
&  \left.  \times e^{-ikx}+\left[  a_{1x^{\prime}}^{\ast}\left(
\mathbf{k}\right)  -a_{-1x^{\prime}}^{\ast}\left(  \mathbf{k}\right)  \right]
e^{ikx}\right\}  \label{A}%
\end{align}
where%
\begin{equation}
a_{rx^{\prime}}\left(  \mathbf{k}\right)  =a_{r}\left(  \mathbf{k}\right)
e\left(  \mathbf{k}\right)  e^{ikx^{\prime}}, \label{a}%
\end{equation}
the subscript $-1$ denotes a series that is odd under exchange of photons and
antiphotons so that $A\rightarrow-A$, the subscript $1$ denotes an even series
so $A\rightarrow A$ and $a_{r}\left(  \mathbf{k}\right)  $ for $r=\pm1$ a real
Lorentz scalar. If $a_{1}=0$ this is an odd series, if $a_{-1}=0$ it is an
even series so
\begin{equation}
A\left(  x\right)  \mathbf{=}A_{1}\left(  x\right)  +iA_{-1}\left(  x\right)
\label{Aeo}%
\end{equation}
where $A_{1}$ and $A_{-1}$ are real. If $a_{1}=a_{-1}$, $A\left(  x\right)  $
is a positive energy photon term, while if $a_{-1}=-a_{1}$, it is a negative
energy antiphoton term. The four-vector $x^{\prime}$ defines the spacetime
origin. The real four-vectors $A_{1}$ and $A_{-1}$ in (\ref{Aeo}) replace
$A_{c}$ and $A_{s}$ in \cite{RealHilbert} to allow for rotation of CP light
that mixes the sine and cosine terms. The subscript $t$ on the integral
denotes evaluation at a fixed time $t$ and $d\mathbf{k}\equiv d^{3}k$ is an
infinitesimal volume in $\mathbf{k}$-space. The above form was selected
because lim$_{V\rightarrow\infty}\Delta\mathbf{n}/V=d\mathbf{k}/\left(
2\pi\right)  ^{3}$ where $\Delta\mathbf{n}$ is the number of states and $\int
d^{4}k\delta\left(  \omega_{k}^{2}/c^{2}-\left\vert \mathbf{k}\right\vert
^{2}\right)  =\int_{t}\frac{d\mathbf{k}}{2\omega_{k}/c}$ is invariant. The
electric and magnetic fields are
\begin{equation}
\mathbf{E}\left(  x\right)  =-\partial_{t}\mathbf{A}\left(  x\right)
-\mathbf{\nabla}\phi\left(  x\right)  ,\ \mathbf{B}\left(  x\right)
=\mathbf{\nabla}\times\mathbf{A}\left(  x\right)  . \label{1photonEandB}%
\end{equation}
The Mostafazadeh sign of energy operator that is useful for application to
linear combinations of positive and negative energy fields is
\cite{MostafazadehZamani}
\begin{equation}
\widehat{\epsilon}\equiv i\left(  -\nabla^{2}\right)  ^{-1/2}\partial_{ct}.
\label{epsilon}%
\end{equation}
In this expression the operator $\left(  -\nabla^{2}\right)  ^{-1/2}$ extracts
a factor $\left\vert \mathbf{k}^{2}\right\vert ^{-1/2}$ from the plane wave
$e^{-i\epsilon kx}$, while $i\partial_{t}e^{-i\epsilon kx}=\epsilon\omega
_{k}e^{-i\epsilon kx}$ so that the operator $\widehat{\epsilon}$ gives the
sign of energy, $\epsilon=\pm$.

The Lagrangian describing the real fields $A_{1}$ and $A_{-1}$ can be written
in the complex form (\ref{Aeo}) provided this field and its complex conjugate
are treated as formally independent \cite{CT}. The standard Lagrangian density
$\mathcal{L}=\epsilon_{0}\left(  \mathbf{E\cdot E}^{\ast}-c^{2}\mathbf{B\cdot
B}^{\ast}\right)  \mathcal{-}J_{m}^{\mu\ast}A_{\mu}-J_{m}^{\mu}A_{\mu}^{\ast}%
$, with matter four-current $J_{m}$ pure imaginary since it is odd, gives the
classical Maxwell equations and conservation laws for energy, momentum and
total angular momentum. In the Coulomb gauge in which $\mathbf{A=A}_{\perp}$
is transverse, the canonical momentum conjugate to $\mathbf{A}_{\perp}$ is
$-\epsilon_{0}\mathbf{E}_{\perp}^{\ast}$, the momentum conjugate to
$\mathbf{A}_{\perp}^{\ast}$ is $-\epsilon_{0}\mathbf{E}_{\perp}$, and the
conserved density generated by a global phase change is $-\epsilon_{0}\left(
\mathbf{E}_{\perp}^{\ast}\mathbf{\cdot A}_{\perp}\mathbf{-E}_{\perp
}\mathbf{\cdot A}_{\perp}^{\ast}\right)  =2\epsilon_{0}\mathbf{E}_{\perp
}\mathbf{\cdot A}_{\perp}^{\ast}$. If the Coulomb gauge is specified the
subscript $\perp$ is redundant, but it is retained here since the transverse
part of $A$ is gauge independent. Writing the transverse part of (\ref{A}) as%
\begin{align}
\mathbf{A}_{\perp}\left(  x\right)   &  =\sqrt{\frac{\hbar}{2\epsilon_{0}}%
}\sum_{\lambda=\pm1}\int_{t}\frac{d\mathbf{k}}{\left(  2\pi\right)  ^{3}%
\omega_{k}}\left[  \mathbf{a}_{\lambda+}\left(  \mathbf{k}\right)
e^{-ikx}\right.  \nonumber\\
&  \left.  +\mathbf{a}_{\lambda-}\left(  \mathbf{k}\right)  e^{ikx}\right]
\label{Atrans}%
\end{align}
for brevity and evaluating $\mathbf{E}_{\perp}\left(  x\right)  =-\partial
_{t}\mathbf{A}_{\perp}\left(  x\right)  $, the gauge invariant conserved
quantity becomes
\begin{equation}
2\epsilon_{0}\int d\mathbf{xE}_{\perp}\mathbf{\cdot A}_{\perp}^{\ast}%
=\sum_{\lambda=\pm1}\int_{t}\frac{d\mathbf{k}}{\left(  2\pi\right)  ^{3}%
}\left[  \left\vert \mathbf{a}_{\lambda+}\left(  \mathbf{k}\right)
\right\vert ^{2}-\left\vert \mathbf{a}_{\lambda-}\left(  \mathbf{k}\right)
\right\vert ^{2}\right]  .\label{conserved}%
\end{equation}
For the real potentials $A_{1}\left(  x\right)  $ and $A_{-1}\left(  x\right)
$ this equals zero. Creation or annihilation of photon/antiphotons pairs is
consistent with this conservation law, while creation or annihilation of
unaccompanied positive frequency photons or negative frequency antiphotons
violates it.

The interpretation of (\ref{conserved}) as a conservation law is new but its
form motivated the definition of scalar product used in \cite{RealHilbert} and
previous work. If $\mathbf{E}_{\perp}$ is replaced with $\widehat{\epsilon
}\mathbf{E_{\perp}\equiv}\widetilde{\mathbf{E}}_{\perp}$ so that the sign of
the antiphoton terms is changed, the positive definite number density%

\begin{equation}
\rho\left(  x\right)  =\frac{\epsilon_{0}}{\hbar}\widetilde{\mathbf{E}}%
_{\perp}^{\ast}\left(  x\right)  \mathbf{\cdot A}_{\perp}\left(  x\right)
\label{rho}%
\end{equation}
is obtained. Details of the contribution of the longitudinal and scalar
components to the number density and scalar product will not be presented here
but in \cite{MaxwellQM} it was found that in the Coulomb gauge only transverse
waves propagate while in the Lorenz gauge the contributions of longitudinal
and scalar photons to number density cancel. As in{{ \cite{RealHilbert}}}
t{{he scalar product}} of states $A_{1}$ and $A_{2}${{ will be defined as}}%
\begin{equation}
\left(  A_{1},A_{2}\right)  _{t}=\frac{\epsilon_{0}}{\hbar}\sum_{\lambda=\pm
}\int_{t}d\mathbf{x}\widetilde{\mathbf{E}}_{1\lambda}^{\ast}\left(  x\right)
\cdot\mathbf{A}_{2\lambda}\left(  x\right)  . \label{ScalarProduct}%
\end{equation}
In bra-ket notation with substitution of (\ref{A}) and use of the
Parseval-Plancherel identity the $r=\pm1$, $\lambda=\pm1$ terms of
(\ref{ScalarProduct}) can be written as
\begin{align}
\left\langle \widetilde{\mathbf{E}}_{1r\lambda}\cdot\mathbf{A}_{2r^{\prime
}\lambda^{\prime}}\right\rangle  &  =\left\langle \widetilde{E}_{1r\lambda
}|A_{2r^{\prime}\lambda}\right\rangle \delta_{\lambda\lambda^{\prime}}%
\delta_{rr^{\prime}}\label{BraKet}\\
&  =\int\frac{d\mathbf{k}}{\left(  2\pi\right)  ^{3}\omega_{k}}a_{1r\lambda
}\left(  \mathbf{k}\right)  a_{2r^{\prime}\lambda^{\prime}}\left(
\mathbf{k}\right) \nonumber\\
&  \times e^{ik\left(  x_{1}-x_{2}\right)  }\delta_{\lambda\lambda^{\prime}%
}\delta_{rr^{\prime}}. \label{kscalarproduct}%
\end{align}
where $A_{jr\lambda}\equiv\left\vert \mathbf{A}_{jr\lambda}\right\vert $ and
$\widetilde{E}_{jr\lambda}\equiv\left\vert \widetilde{\mathbf{E}}_{jr\lambda
}\right\vert $. Inspection of (\ref{ScalarProduct}) to (\ref{kscalarproduct})
shows that these expressions for the scalar product involve both the vector
potential and the electric field rather than a single function. QM based on
scalar products of this form can be described within the formalism of
biorthogonal QM \cite{Brody,HawtonDebierre,Beige}.

The one-photon Hilbert space will be defined as the space of all
four-potentials of the form (\ref{A}) with scalar product (\ref{ScalarProduct}%
) to (\ref{kscalarproduct}). Eigenvectors of observables will be written in
positive energy form in real space so that the basis includes both even and
odd fields. It can be verified by substitution in (\ref{kscalarproduct}) that
the transverse plane waves with definite momenta $\hbar\mathbf{k}^{\prime}$
and helicity $\lambda^{\prime}$ defined covariantly as%
\begin{equation}
\mathbf{a}_{r^{\prime}\lambda^{\prime}\mathbf{k}^{\prime}}\left(
\mathbf{k}\right)  =\left(  2\pi\right)  ^{3}\omega_{k}\delta\left(
\mathbf{k}-\mathbf{k}^{\prime}\right)  \mathbf{e}_{\lambda^{\prime}}\left(
\mathbf{k}\right)  \label{PlaneWaves}%
\end{equation}
for $r^{\prime}=\pm1$ with $\omega_{k}=c\left\vert \mathbf{k}\right\vert $ are
biorthogonal in the sense that $\left(  A_{r\lambda\mathbf{k}},A_{r^{\prime
}\lambda^{\prime}\mathbf{k}^{\prime}}\right)  =\delta_{rr^{\prime}}%
\delta_{\lambda\lambda^{\prime}}\left(  2\pi\right)  ^{3}\omega_{k}%
\delta\left(  \mathbf{k}-\mathbf{k}^{\prime}\right)  .$ In position space
substitution in (\ref{A}) gives $\mathbf{A}_{\lambda^{\prime}\mathbf{k}%
^{\prime}}\left(  \mathbf{x}\right)  =\sqrt{\frac{2\hbar}{\epsilon_{0}}%
}e^{i\mathbf{k}^{\prime}\cdot\mathbf{x}}\mathbf{e}_{\lambda^{\prime}}\left(
\mathbf{k}^{\prime}\right)  .$ Position is also an observable. The Fourier
transform of the localized state $\delta\left(  \mathbf{x}-\mathbf{x}^{\prime
}\right)  $ at $\mathbf{x}^{\prime}$ is the plane wave $\exp\left(
-i\mathbf{k\cdot x}^{\prime}\right)  $ so the photon position eigenvectors in
the Schr\"{o}dinger picture (SP) should be of the form (\ref{a}) with
\begin{equation}
\mathbf{a}_{r\lambda^{\prime}\mathbf{x}^{\prime}}\left(  \mathbf{k}\right)
=\mathbf{e}_{\lambda^{\prime}}\left(  \mathbf{k}\right)  e^{-i\mathbf{k}%
\cdot\mathbf{x}^{\prime}},\ a_{r\lambda^{\prime}x^{\prime}}\left(
\mathbf{k}\right)  =1 \label{x}%
\end{equation}
for $r=\pm1$.

Pulse propagation takes place in real space. The projection of an arbitrary
physical state of the form (\ref{a}) onto the $A_{x\lambda}$ basis, evaluated
using (\ref{kscalarproduct}),%
\begin{equation}
\phi_{r\lambda}\left(  x\right)  =\left(  A_{x\lambda},A_{r}\right)  =\int%
_{t}\frac{d\mathbf{k}}{\left(  2\pi\right)  ^{3}\omega_{k}}a_{r\lambda}\left(
\mathbf{k}\right)  e^{-ikx}, \label{phi}%
\end{equation}
is the probability amplitude for the photon in state $\left\vert
A_{r}\right\rangle $ to be at $\mathbf{x}$ at time $t$. Setting $a_{r\lambda
}\left(  \mathbf{k}\right)  $ in (\ref{phi}) equal to $a_{r\lambda x^{\prime}%
}\left(  \mathbf{k}\right)  $ with $\Delta t\equiv t-t^{\prime}$ and
$R\equiv\left\vert \mathbf{x}-\mathbf{x}^{\prime}\right\vert $ an explicit
expression for its time evolution can be obtained by taking sums and
differences of%
\begin{align}
\int_{t}\frac{d\mathbf{k}}{\left(  2\pi\right)  ^{3}\omega_{k}}e^{-ik\left(
x-x^{\prime}\right)  }  &  =\frac{1}{4\pi^{2}r}\sum_{\gamma=\pm}\left[
i\gamma\pi\delta\left(  R-\gamma c\Delta t\right)  \right. \label{I}\\
&  \left.  +P\left(  \frac{1}{R-\gamma c\Delta t}\right)  \right] \nonumber
\end{align}
where $P$ is the principal value and the sum over $\gamma$ comes from
integration over the $\mathbf{k}$-space polar angle and represents a sum over
incoming and outgoing spherical waves. For the even field $A_{1}$ ,
$\phi_{1\lambda}\left(  x\right)  =\left(  A_{x\lambda},A_{1x^{\prime}\lambda
}\right)  $ is nonlocal and, since $J_{m}$ is odd and hence pure imaginary,
there is no source term in its equation of motion. It is completely decoupled
from charge matter and thus, if it exists at all, cannot be detected. Only the
imaginary odd term in (\ref{I})%
\begin{align}
\phi_{-1\lambda x^{\prime}}\left(  x\right)   &  =\left(  A_{x\lambda
},A_{-1x^{\prime}\lambda}\right) \nonumber\\
&  =\frac{1}{4\pi r}\left[  \delta\left(  R+c\Delta t\right)  -\delta\left(
R-c\Delta t\right)  \right]  . \label{phio}%
\end{align}
couples to charged matter. In a source free region there is no absorption or
emission and the photon just passes through $\mathbf{x}^{\prime}$ at time
$t^{\prime}$.

Expression (\ref{phio}) satisfies the homogeneous Klein Gordon (KG) equation
and equals the advanced minus the retarded potential. The retarded potential
is important in classical EM and (\ref{phio}) shows that it can be calculated
for one-photon states. In the presence of a source such as an atom or a
quantum dot the wave equation describing propagating transverse photons is
\begin{equation}
\square\phi_{-1\lambda}\left(  x\right)  =J_{m\lambda}\left(  x\right)
\label{WaveEq}%
\end{equation}
where $J_{m\lambda}$ is the $\lambda$ component of the matter four-current.
The general solution to this wave equation is a particular solution determined
by $J_{m\lambda}$ plus a general solution to the homogeneous wave equation
$\square\phi_{-1\lambda}\left(  x\right)  =0$. Schweber \cite{Schweber}
inverted $\square$ and found that the unique Green's function solving $\square
G_{\lambda x^{\prime}}\left(  x\right)  =\delta\left(  \mathbf{x}%
-\mathbf{x}^{\prime}\right)  \delta\left(  t-t^{\prime}\right)  $ is
\begin{equation}
G_{\lambda x^{\prime}}\left(  x\right)  =\frac{1}{4\pi R}\left[  \delta\left(
R+c\Delta t\right)  +\delta\left(  R-c\Delta t\right)  \right]  \label{Green}%
\end{equation}
where $t^{\prime}=t-R/c<t$ is the retarded time and $t^{\prime}=t+R/c>t$ is
the advanced time. He concluded that the retarded potential is determined by
boundary conditions. The particular solution to (\ref{WaveEq}) for a source of
helicity $\lambda$ is \cite{Schweber}
\begin{align}
\phi_{-1\lambda}^{\left(  p\right)  }\left(  x\right)   &  =\int%
\frac{d\mathbf{x}^{\prime}}{4\pi r}\left[  H\left(  \Delta t-R/c\right)
J_{m\lambda}\left(  \mathbf{x}^{\prime},\Delta t-R/c\right)  \right.
\nonumber\\
&  \left.  +H\left(  -\Delta t-R/c\right)  J_{m\lambda}\left(  \mathbf{x}%
^{\prime},\Delta t+R/c\right)  \right]  \label{Particular}%
\end{align}
where $H\left(  s\right)  =0$ for $s<0$ and $1$ for $s\geq0$ is the Heaviside
step function. The one photon probabiity amplitude emitted instantaneously at
$t^{\prime}$ by a localized source at $\mathbf{x}^{\prime}$ is $\frac{1}%
{2}\left[  G_{\lambda x^{\prime}}\left(  x\right)  -\phi_{-1\lambda x^{\prime
}}\left(  x\right)  \right]  $ given by (\ref{phio}) and (\ref{Green}) while
$\frac{1}{2}\left[  G_{\lambda x^{\prime}}\left(  x\right)  +\phi_{-1\lambda
x^{\prime}}\left(  x\right)  \right]  $ describes absorption. Eq.
(\ref{Particular}) in combination with a generalization of (\ref{phio}) based
on (\ref{phi}) can be used to describe emission or absorption by a more
realistic one photon source.

The potential $\phi_{-1\lambda x^{\prime}}\left(  x\right)  $ given by
(\ref{phio}) is a Lorentz scalar whose time derivative, $i\partial_{t}%
\phi_{-1\lambda x^{\prime}}\left(  x\right)  $, is a density. At $t=t^{\prime
}$,%
\begin{equation}
\psi_{-1\lambda\mathbf{x}^{\prime}}\left(  t,\mathbf{x}\right)  =\int_{t}%
\frac{d\mathbf{k}}{\left(  2\pi\right)  ^{3}}e^{-i\mathbf{k\cdot}\left(
\mathbf{x}-\mathbf{x}^{\prime}\right)  }=\delta\left(  \mathbf{x}%
-\mathbf{x}^{\prime}\right)  \label{x_basis}%
\end{equation}
forms a localized basis. The Born rule gives a probability interpretation of
the state vector. Expanding $\psi_{-1\lambda}$ in the $\delta$-basis at time
$t$ as%
\begin{equation}
\psi_{-1\lambda}\left(  t,\mathbf{x}\right)  =\int_{t}d\mathbf{x}^{\prime
}\delta\left(  \mathbf{x}-\mathbf{x}^{\prime}\right)  \psi_{-1\lambda}\left(
t,\mathbf{x}^{\prime}\right)  , \label{pbasis}%
\end{equation}
it can be seen that $\psi_{-1\lambda}\left(  x\right)  $ is the probability
amplitude for a photon to be in the state $\delta\left(  \mathbf{x}%
-\mathbf{x}^{\prime}\right)  $ on the $t$-hyperplane.The $\lambda$-helicity
$\mathbf{x}$-space and $\mathbf{k}$-space probability densities are
\begin{align}
\rho_{-1\lambda}\left(  t,\mathbf{x}\right)   &  =\left\vert \psi_{-1\lambda
}\left(  t,\mathbf{x}\right)  \right\vert ^{2},\label{xdensity}\\
\rho_{-1\lambda}\left(  \mathbf{k}\right)   &  =\left\vert a_{-1\lambda
}\left(  \mathbf{k}\right)  \right\vert ^{2}. \label{kdensity}%
\end{align}
where%
\begin{equation}
\sum_{\lambda=\pm1}\int d\mathbf{x}\left\vert \psi_{-1\lambda}\left(
t,\mathbf{x}\right)  \right\vert ^{2}=\sum_{\lambda=\pm1}\int\frac
{d\mathbf{k}}{\left(  2\pi\right)  ^{3}}\left\vert a_{-1\lambda}\left(
\mathbf{k}\right)  \right\vert ^{2}=1 \label{normalized}%
\end{equation}
for normalized physical states.

\section{QED Fock space and multiphoton states}

The complete photon state space is determined by QED. Photons are bosons so
there is no exclusion principle and $n$-photon states are allowed for $n=0$ or
any positive integral value of $n$. In a plane wave basis the covariant photon
commutation relations are
\begin{align}
\left[  \widehat{a}_{\lambda}\left(  \mathbf{k}\right)  ,\widehat{a}%
_{\lambda^{\prime}}\left(  \mathbf{k}^{\prime}\right)  \right]   &
=0,\ \left[  \widehat{a}_{\lambda}^{\dagger}\left(  \mathbf{k}\right)
,\widehat{a}_{\lambda^{\prime}}^{\dagger}\left(  \mathbf{k}^{\prime}\right)
\right]  =0,\nonumber\\
\left[  \widehat{a}_{\lambda}\left(  \mathbf{k}\right)  ,\widehat{a}%
_{\lambda^{\prime}}^{\dagger}\left(  \mathbf{k}^{\prime}\right)  \right]   &
=\delta_{\lambda,\lambda^{\prime}}\left(  2\pi\right)  ^{3}\omega_{k}%
\delta\left(  \mathbf{k}-\mathbf{k}^{\prime}\right)  . \label{kcommutation}%
\end{align}
where the operator $\widehat{a}_{\lambda}\left(  \mathbf{k}\right)  $
annihilates a photon with wave vector $\mathbf{k}$ and helicity $\lambda$ and
$\widehat{a}_{\lambda}^{\dagger}\left(  \mathbf{k}\right)  $ creates one.
Using these commutation relations it can be verified that%
\begin{equation}
\left\vert a_{\lambda\mathbf{k}n}\right\rangle =\frac{\left[  \widehat{a}%
_{\lambda n}^{\dag}\left(  \mathbf{k}\right)  \right]  ^{n}}{\sqrt{n!}%
}\left\vert 0\right\rangle \label{nphotonplanewave}%
\end{equation}
are normalized $n$-photon states where $\left\vert 0\right\rangle $ is the
zero-photon (vacuum) state.

The field operators can be obtained by second quantization of any real field
so we choose the odd field,%
\begin{align}
\widehat{\mathbf{A}}\left(  x\right)   &  =-i\sqrt{\frac{\hbar}{\epsilon_{0}}%
}\sum_{\lambda=\pm1}\int_{t}\frac{d\mathbf{k}}{\left(  2\pi\right)  ^{3}%
\omega_{\mathbf{k}}}\left[  \widehat{a}_{\lambda}\left(  \mathbf{k}\right)
\mathbf{e}_{\lambda^{\prime}}\left(  \mathbf{k}\right)  e^{-ikx}\right.
\nonumber\\
&  \left.  -\widehat{a}_{\lambda}^{\dagger}\left(  \mathbf{k}\right)
\mathbf{e}_{\lambda^{\prime}}^{\ast}\left(  \mathbf{k}\right)  e^{ikx}\right]
. \label{Aop}%
\end{align}
The transverse electric and magnetic field operators can then be obtained by
differentiation as
\begin{equation}
\widehat{\mathbf{E}}_{\perp}\left(  x\right)  =-\partial_{t}%
\widehat{\mathbf{A}}\left(  x\right)  ,\ \widehat{\mathbf{B}}\left(  x\right)
=\mathbf{\nabla}\times\widehat{\mathbf{A}}\left(  x\right)  . \label{EandBops}%
\end{equation}
In QED causality is enforced by the commutation relations. Defining%
\begin{align}
\widehat{C}_{\lambda}\left(  x,x^{\prime}\right)   &  \equiv i\frac
{\epsilon_{0}}{\hbar}\left[  \widehat{\mathbf{A}}_{\lambda}\left(
t,\mathbf{x}\right)  \cdot\widehat{\mathbf{E}}_{\lambda}\left(  t^{\prime
},\mathbf{x}^{\prime}\right)  \right. \nonumber\\
&  \left.  -\widehat{\mathbf{E}}_{\lambda}\left(  t^{\prime},\mathbf{x}%
^{\prime}\right)  \cdot\widehat{\mathbf{A}}_{\lambda}\left(  t,\mathbf{x}%
\right)  \right]  \label{C}%
\end{align}
it can be verified by substitution at $t=t^{\prime}$ that $\widehat{C}%
_{\lambda}\left(  t,\mathbf{x};t,\mathbf{x}^{\prime}\right)  =\delta\left(
\mathbf{x}-\mathbf{x}^{\prime}\right)  $ and
\begin{equation}
\left\langle 0\left\vert \widehat{C}_{\lambda}\left(  x,x^{\prime}\right)
\right\vert 0\right\rangle =\phi_{-1\lambda}\left(  x\right)  .
\label{VacuumExpectation}%
\end{equation}
Thus the causal propagation described by (\ref{phio}) is consistent with QED
where in QED the sign change of the antiphoton term is a consequence of the
bosonic commutation relations.\ 

The QED positive and negative energy annihilation and creation opertors define
the plane wave and localized bases but they do no extend to the description of
real photon fields in an obvious way. I found first quantization to be more
convenient for this purpose. First quantized one photon states were the
subject of the previous Section and, following the rules of QM for bosons,
multiphoton states can be written as symmetrized products of one-photon
states. For photons at $\mathbf{x}_{1}$ and $\mathbf{x}_{2}$ in odd states
$A_{-1j}$ and $A_{-1k}$ the symmetrized two-photon state is%
\begin{align}
A_{jk}\left(  \mathbf{x}_{1},\mathbf{x}_{2},t\right)   &  =\frac{1}{\sqrt{2}%
}\left[  A_{-1j}\left(  \mathbf{x}_{1},t\right)  A_{-1k}\left(  \mathbf{x}%
_{2},t\right)  \right. \nonumber\\
&  \left.  +A_{-1j}\left(  \mathbf{x}_{2},t\right)  A_{-1k}\left(
\mathbf{x}_{1},t\right)  \right]  \label{twophoton}%
\end{align}
at time $t$. This ensures that the scalar product $\left(  A_{4}A_{3}%
,A_{2}A_{1}\right)  =\left(  A_{4},A_{2}\right)  \left(  A_{3},A_{1}\right)
+\left(  A_{4},A_{1}\right)  \left(  A_{3},A_{2}\right)  $ does not depend on
photon order in the two-photon state (\ref{twophoton}). If both photons are in
the same state, $A_{jj}\left(  \mathbf{x}_{1},\mathbf{x}_{2},t\right)
=A_{-j}\left(  \mathbf{x}_{1},t\right)  A_{-j}\left(  \mathbf{x}_{2},t\right)
$ is symmetric.

As an example we consider photon pulses travelling in the $+$ or $-$ direction
in a one-dimensional wave guide, $A_{-1\lambda}(x\pm ct).$ For a photon
propagating in one dimension states with definite helicity $\lambda$ are just
circularly polarized. A single-photon passed through a beam splitter at $x=0$
described by%
\begin{equation}
A_{-1\lambda}\left(  x,t\right)  =\frac{1}{\sqrt{2}}\left[  A_{-1\lambda
}(x-ct)+A_{-1\lambda}(x+ct)\right]  \label{split}%
\end{equation}
is equally likely to be counted on the positive or negative $x$-axis. For
entangled photons with total linear and angular momentum zero, perhaps created
by position annihilation at $x=0$, the two-photon state%
\begin{align}
A\left(  x_{1},x_{2},t\right)   &  =\frac{1}{\sqrt{2}}\sum_{\lambda=\pm
1}\left[  A_{-1\lambda}(x_{1}-ct)A_{-1-\lambda}(x_{2}+ct)\right. \nonumber\\
&  \left.  +A_{-1-\lambda}(x_{1}-ct)A_{-1\lambda}(x_{2}+ct)\right]
\label{entangled}%
\end{align}
is symmetrized under exchange of photons at $\mathbf{x}_{1}$ and
$\mathbf{x}_{2}$ by the sum over $\lambda$. Detection of a photon with
helicity $\lambda$ at $x_{1}=ct$ collapses the entangled state
(\ref{entangled}) to $A_{-1-\lambda}(x_{1}+ct)$.

A coherent state with helicity $\lambda$, wave vector $\mathbf{k}$ and average
photon number $\alpha_{\lambda}^{\ast}\alpha_{\lambda}$ is $\left\vert
\alpha_{\lambda\mathbf{k}}\right\rangle =\sum_{n=0}^{\infty}\alpha
_{\mathbf{k}\lambda}^{n}\left\vert a_{\lambda\mathbf{k}n}\right\rangle $ where
$\widehat{a}_{\lambda}\left(  \mathbf{k}\right)  \left\vert \alpha
_{\mathbf{k}\lambda}\right\rangle =\alpha_{\lambda\mathbf{k}}\left\vert
\alpha_{\lambda\mathbf{k}}\right\rangle $ and $\left\langle \alpha
_{\lambda\mathbf{k}}\right\vert \widehat{a}_{\lambda}^{\dagger}\left(
\mathbf{k}\right)  =\alpha_{\lambda\mathbf{k}}^{\ast}\left\langle
\alpha_{\lambda\mathbf{k}}\right\vert $. If the plane waves (\ref{PlaneWaves})
are in coherent states $\left\vert \left\{  \alpha_{\mathbf{k}\lambda
}\right\}  \right\rangle $ for all basis state $\left(  \mathbf{k,\lambda
}\right)  $ the expectation values of the vector potential operator is%
\begin{align}
\mathbf{A}_{cl}  &  \equiv\left\langle \left\{  \alpha_{\lambda\mathbf{k}%
}\right\}  \left\vert \widehat{\mathbf{A}}\left(  x\right)  \right\vert
\left\{  \alpha_{\lambda\mathbf{k}}\right\}  \right\rangle \nonumber\\
&  =i\sqrt{\frac{\hbar}{\epsilon_{0}}}\sum_{\lambda=\pm1}\int_{t}%
\frac{d\mathbf{k}}{\left(  2\pi\right)  ^{3}\omega_{k}}\left[  \alpha
_{\lambda\mathbf{k}}\mathbf{e}_{\lambda^{\prime}}e^{-ikx}\right. \nonumber\\
&  \left.  -\alpha_{\lambda\mathbf{k}}^{\ast}\left(  \mathbf{k}\right)
\mathbf{e}_{\lambda^{\prime}}^{\ast}\left(  \mathbf{k}\right)  e^{ikx}\right]
. \label{expectation}%
\end{align}
After taking expectation values of the operator (\ref{Aop}) to get
(\ref{expectation}) detailed information about the distribution over photon
number is lost and only its average value is retained. The averaging process
just counts transitions between states that differ by a photon number of $1$.
For states containing a definite number of photons, in particular for
one-photon states, expectation values of the field operators are $0$.

\section{Summary and Conclusion{}}

Photon quantum mechanics as described here preserves the classical form of the
EM potential and fields when first and second quantized. Only the
interpretation need be changed - from real observable classical fields, to
probability amplitudes, and then to operators that create and annihilate
photons. The real potentials are even and odd under QFT charge conjugation,
but only those that are odd couple to charged matter and can be localized in a
finite region. Here even and odd fields were written as the real and imaginary
parts of a complex field whose use simplifies the mathematics and facilitates
use of the standard Lagrangian and relativistic scalar product. The number
density if $\left(  \epsilon_{0}/\hbar\right)  \widetilde{\mathbf{E}}\left(
x\right)  \cdot\mathbf{A}\left(  x\right)  $ where in $\widetilde{\mathbf{E}}$
the sign of the antiphoton term is changed analogous to the effect of the
$\beta$ matrix in the Dirac theory of electrons and positrons and the scalar
product is based on its spatial integral.\textbf{ }Eqs. (\ref{phio}) to
(\ref{expectation}) provide a new scalar description of single photon states
with a well defined physical interpretation that may prove to be useful in applications.

In conclusion, one photon fields of the classical form combined with a
probabilistic interpretation can be used in quantum optics, quantum computing
and quantum information without loss of rigor. This can be extended to
multiphoton states, possibly entangled, by construction of a symmetrized
product of one-photon states. In the opinion of this author, this conclusion
has important practical and fundamental implications.

\end{document}